\documentclass[conference]{IEEEtran}
\usepackage{amsmath,bm,amsfonts,amssymb,graphicx,color,mathtools,setspace,
comment,multirow,xfrac}
\usepackage{booktabs}
\usepackage{epstopdf}
\usepackage{mathtools}
\usepackage{cite,citesort}

\newcommand*{\QED}{\hfill\ensuremath{\blacksquare}}

\begin{document}
\title{Uplink Analysis of Large MU-MIMO Systems With Space-Constrained 
Arrays in Ricean Fading}
 \author{\IEEEauthorblockA{Harsh Tataria\IEEEauthorrefmark{1},
  											Peter J. Smith\IEEEauthorrefmark{2},
  											Michail Matthaiou\IEEEauthorrefmark{3}, and
  											Pawel A. Dmochowski\IEEEauthorrefmark{1}
  											}
  \IEEEauthorblockA{\IEEEauthorrefmark{1}
  \small{School of Engineering and Computer Science, Victoria 
 University of Wellington, Wellington, New Zealand}}
  \IEEEauthorblockA{\IEEEauthorrefmark{2}
  \small{School of Mathematics and Statistics, Victoria 
 University of Wellington, Wellington, New Zealand}}
  \IEEEauthorblockA{\IEEEauthorrefmark{3}
  \small{School of Electronics, Electrical Engineering and Computer Science, 
  Queen's University Belfast, Belfast, Northern Ireland, UK}}
  \IEEEauthorblockA{\small{email:\{harsh.tataria, pawel.dmochowski\}@ecs.vuw.ac.nz,~peter.smith@vuw.ac.nz,~m.matthaiou@qub.ac.uk}}}

% make the title area
\maketitle

\begin{abstract}
Closed-form approximations to the 
expected per-terminal signal-to-interference-plus-noise-ratio (SINR) and 
ergodic sum spectral efficiency of a large multiuser multiple-input 
multiple-output system are presented. Our analysis assumes correlated Ricean 
fading with maximum ratio combining on the uplink, where the base station 
(BS) is equipped with a uniform linear array (ULA) with physical size 
restrictions. Unlike previous studies, our model caters for the presence of 
unequal correlation matrices and unequal Rice factors for each 
terminal. As the number of BS antennas grows without bound, with a finite 
number of terminals, we derive the limiting expected per-terminal 
SINR and ergodic sum spectral efficiency of the system. Our findings suggest 
that with restrictions on the size of the ULA, the expected SINR saturates 
with increasing operating signal-to-noise-ratio (SNR) and BS antennas. 
Whilst unequal correlation matrices result in higher performance, 
the presence of strong line-of-sight (LoS) has an opposite effect. 
Our analysis accommodates changes in system dimensions, SNR, LoS levels, 
spatial correlation levels and variations in 
fixed physical spacings of the BS array. 
\end{abstract}
\IEEEpeerreviewmaketitle

\vspace{-8pt}
\section{Introduction}
\label{Introduction}
The deployment of large numbers of antennas at a cellular 
base station (BS) to communicate with multiple user terminals 
has received a considerable amount of 
attention recently \cite{RUSEK,LARSSON}. Specifically, large (a.k.a. massive) 
multiuser multiple-input multiple-output (MU-MIMO) systems have been 
shown to achieve orders of magnitude greater performance than conventional 
MU-MIMO systems, due to their ability to leverage 
favorable propagation conditions \cite{LARSSON}. 
Nevertheless, the emergence of such 
systems has posed new engineering challenges which must be overcome 
before their adoption on a scale commensurate with their true potential. 
One of the critical issues is accommodating large numbers of antennas in
fixed physical spacings \cite{MASOUROS1,MASOUROS2}. This tends to  
increase the level of spatial correlation and antenna coupling, as 
successive elements are placed in close proximity with inter-element 
spacings less than the desired half-a-wavelength \cite{MASOUROS1}. 
This is known to cause a detrimental impact on the terminal 
signal-to-interference-plus-noise-ratio (SINR) and system spectral 
efficiency. It is thus important to rigorously 
analyze and evaluate the performance of systems with space-constrained (SC) 
antenna arrays. 

Numerous works have investigated the impact of SC
antenna arrays on the performance of large MU-MIMO systems (see 
e.g., \cite{MASOUROS1,MASOUROS2,NGO,GARCIA-ROD,ZHANG,BISWAS,GE} and 
references therein). Specifically, \cite{MASOUROS1} analyzed the 
ergodic sum spectral efficiency of large MU-MIMO systems with 
fixed array dimensions. The authors in \cite{MASOUROS2} demonstrated that 
multiuser interference does not vanish in SC MU-MIMO systems with 
growing numbers of antennas. The uplink performance 
with maximum-ratio combining (MRC), zero-forcing and 
minimum-mean-squared-error receivers has been analyzed in 
\cite{NGO,ZHANG} where the authors derive upper and lower bounds on the 
ergodic sum spectral efficiency. Moreover, 
\cite{GARCIA-ROD,BISWAS,GE} investigated the 
energy efficiency performance of SC systems with various 
large-scale antenna array topologies considering antenna coupling. 

However, very few of the above mentioned studies\footnote{We make an 
exception in \cite{MASOUROS2}, which considers 
pure LoS channels. This is an extreme case, which in general 
may not be realizable in practice, even at mmWave frequencies, where on 
average 1-3 scattering clusters are anticipated in the 
propagation channel (see e.g., \cite{AKDNENIZ}).} consider the effects of 
line-of-sight (LoS) components, which may be a dominant 
feature in future wireless access with the use of smaller cell sizes, potentially 
operating in the millimeter-wave (mmWave) frequency bands 
\cite{SUN,TATARIA,TATARIA2}. Hence, 
understanding the performance of SC systems with LoS presence, 
i.e., with Ricean fading 
is of particular importance. 
Moreover, the respective channel models in \cite{NGO,ZHANG,GE} 
assume that all terminals are seen by the BS array via the same set of 
incident directions, resulting in common (equal) spatial correlation structures. 
In reality, differences in the local scattering around the physical location of 
each terminal gives rise to wide variations in the correlation patterns \cite{NAM}. 
In addition to the small inter-element spacings, 
this further contributes to the level of correlation in the channel, 
impacting the terminal SINR and system spectral efficiency. Thus, 
to more accurately capture the correlation differences in 
multiple channels, we consider distinct correlation matrices for 
each terminal. 
Motivated by the aforementioned considerations, with a SC 
uniform linear array (ULA), we present a 
framework for analyzing the expected per-terminal 
SINR and ergodic sum spectral efficiency of large MU-MIMO 
systems with MRC at the BS. 
Specifically, our main contributions are as follows:
\begin{itemize}
 \item We analyze the performance of MU-MIMO systems with  
 SC ULAs under correlated Ricean 
 fading channels. In doing so, we extend and generalize the 
 SC channel models presented in 
 \cite{MASOUROS1,MASOUROS2,NGO,GARCIA-ROD,ZHANG,BISWAS,GE}
 to cater for unequal correlation matrices and  
 unequal Rice factors for each terminal. 
 To the best of the authors' knowledge, such generality in the 
 channel model has not previously been considered.
 \item With MRC at the BS, we derive tight closed-form 
 approximations to the expected per-terminal SINR and 
 ergodic sum spectral efficiency. We show that a SC 
 antenna deployment causes a saturation of the expected SINR 
 with increasing numbers of BS antennas and operating 
 signal-to-noise-ratios (SNRs). 
 \item With a fixed number of terminals, as the number of BS antennas 
 increases without bound, we derive novel limiting expected 
 SINR and ergodic spectral efficiency 
 expressions to demonstrate the convergence behavior of large 
 SC MU-MIMO systems. 
 \item Finally, we present special cases of the 
 derived analytical results when NLoS components are present with 
 equal and unequal correlation matrices, as well as, when 
 each terminal having LoS has fixed correlation matrices.  
\end{itemize}

\textbf{Notation.} Boldface upper and lower case symbols denote matrices 
and vectors, respectively. Moreover, 
$\bm{I}_{M}$ denotes the $M\times{}M$ identity 
matrix. $\left(\cdot\right)^{\textrm{T}}$, 
$\left(\cdot\right)^{\textrm{H}}$ and $\left(\cdot\right)^{-1}$ denote the 
transpose, Hermitian transpose and inverse operators, respectively. We use 
$\left[\bm{H}\right]_{i,j}$ to refer to the $\left(i,j\right)$-th element of 
$\bm{H}$, whilst $\bm{h}\sim\mathcal{CN}\left(\mu,\sigma^{2}\right)$ denotes 
a complex Gaussian distribution for $\bm{h}$, where each element of $\bm{h}$ 
has a mean $\mu$ and variance $\sigma^{2}$. 
We use $x\sim{}u\left[a,b\right]$ to denote a uniform random variable for 
$x$ taking on values from $a$ to $b$. $||\cdot{}||$, 
$||\cdot{}||_{\textrm{F}}$ and $|\cdot{}|$ denote the standard two norm, 
Frobenius norm and scalar norm, respectively. Finally, 
$\textrm{tr}\left[\cdot\right]$ and $\mathbb{E}\left[\cdot\right]$ denote 
the trace and statistical expectation operations.   

\vspace{-4pt}
\section{System Model}
\label{SystemModel}
\vspace{-1pt}
We consider the uplink of a large MU-MIMO system operating 
in an urban microcellular (UMi) environment, where $L$ non-cooperative 
single-antenna user terminals transmit data to $M$ receive antennas at the BS 
$\left(M\gg{}L\right)$ in the same time-frequency interval. The BS comprises 
of a ULA with equispaced, omnidirectional antennas. 
We assume channel knowledge at the BS with narrow-band transmission and 
no uplink power control. The composite $M\times{}1$ 
received signal at the BS array can be written as 
\vspace{-3pt}
\begin{equation}
 \label{receivedsignal}
 \bm{y}=\rho^{\frac{1}{2}}\bm{G}\bm{D}^{\frac{1}{2}}\bm{s}+\bm{n}, 
 \vspace{-4pt}
\end{equation}
where $\rho$ is the average transmit power of each terminal, $\bm{G}$ denotes 
the $M\times{}L$ fast-fading uplink channel matrix between $M$ BS antennas 
and $L$ terminals (discussed further in Section~\ref{ChannelModel}), 
$\bm{D}$ is an $L\times{}L$ diagonal matrix of link gains for the $L$ terminals 
in the system, such that $\left[\bm{D}\right]_{l,l}=\beta_{l}$. The 
large-scale fading effects are for terminal $l$ are captured 
in $\beta_{l}=\varrho\zeta_{l}\left(r_{0}/r_{l}\right)^{\alpha}$. 
In particular, $\varrho$ denotes the 
unit-less constant for geometric attenuation at a reference distance $r_{0}$, 
$r_{l}$ denotes the link distance between the BS array and terminal $l$, 
$\alpha$ denotes the attenuation exponent and $\zeta_{l}$ models the effects of 
shadow-fading following a log-normal density, i.e., 
$10\log_{10}\left(\zeta_{l}\right)\sim
{}\mathcal{N}\left(0,\sigma_{\textrm{sh}}^{2}\right)$, 
with $\sigma_{\textrm{sh}}$ denoting the shadow-fading standard deviation. 
Numerical values for the above are tabulated in 
Section~\ref{NumericalResults}. 
The $L\times{}1$ vector of uplink data symbols from the $L$ terminals is 
given by $\bm{s}$, such that the $l$-th entry of $\bm{s}$, 
$s_{l}$ has $\mathbb{E}\left[|s_{l}|^{2}\right]=1$. Additive white 
Gaussian noise entries at the $M$ BS antennas is given by the $M\times{}1$ 
vector $\bm{n}$, such that the $l$-th 
entry of $\bm{n}$, $n_{l}\sim{}\mathcal{CN}\left(0,\sigma^{2}\right)$. 
We assume that $\sigma^{2}=1$, hence the average uplink
SNR, defined as $\rho/\sigma^{2}=\rho$.

\vspace{-5pt}
\subsection{Channel Model}
\label{ChannelModel}
\vspace{-3pt}
Previous studies (e.g., \cite{NGO,ZHANG,GE}) on large SC MU-MIMO systems 
consider a physical channel model based on full NLoS propagation conditions, 
where the BS sees the same set of 
scattered directions from each terminal. We extend this model to cater for 
the presence of LoS in the propagation channel, as well as a unique set of scattered 
directions from each terminal taking into account differences in the local scattering 
around each terminal. 
Specifically, $\bm{G}=\left[\bm{g}_{1},\dots{},\bm{g}_{L}\right]$, 
where $\bm{g}_{l}$, the $l$-th column of $\bm{G}$ contains the $M\times{}1$ uplink 
channel vector from terminal $l$ to the BS array given by 
\vspace{-5pt}
\begin{equation}
 \label{terminalgchannel}
 \bm{g}_{l}=\eta'_{l}\hspace{1pt}
 \bm{A}_{l}\hspace{1pt}\bm{h}_{l}\hspace{-1pt}+\hspace{-1pt}
 \bar{\eta}_{l}\hspace{1pt}\bar{\bm{h}}_{l}, 
 \vspace{-5pt}
\end{equation}
where $\eta'_{l}=\eta_{l}\frac{1}{\sqrt{P}}$ with 
$\eta_{l}\hspace{-2pt}=\hspace{-2pt}\big(\frac{1}{1+K_{l}}\big)^{\hspace{-2pt}
\frac{1}{2}}$ and $\bar{\eta_{l}}\hspace{-3pt}=
\hspace{-3pt}\big(\frac{K_{l}}{K_{l}+1}\big)^{\hspace{-2pt}\frac{1}{2}}$. 
In the above, 
$\eta_{l}$ and $\bar{\eta}_{l}$ balance the amount of power present in the diffuse and 
specular components of the channel according to the Ricean $K$-factor, 
$K_{l}$, specific to terminal $l$ \cite{MOLISCH}. Moreover, 
$\eta_{l}$ is further scaled by a factor of $\frac{1}{\sqrt{P}}$ to 
normalize the steering vectors in $\bm{A}_{l}$, the  
$M\times{}P$ receive steering matrix associated with 
the diffuse components of the channel. Here, $P$ denotes a large yet finite 
number of diffuse wavefronts. For ULAs 
\vspace{-4pt}
\begin{equation}
 \label{correlationmatrixterminall}
 \bm{A}_{l}=\left[\bm{a}\left(\phi_{l,1}\right),
 \bm{a}\left(\phi_{l,2}\right),\dots{},\bm{a}\left(\phi_{l,P}\right)\right], 
 \vspace{-3pt}
\end{equation}
where each vector in \eqref{correlationmatrixterminall} is given by 
\vspace{-2pt}
\begin{equation}
 \label{aphili}
 \bm{a}\left(\phi_{l,i}\right)=\left[1,e^{j2\pi{}d\sin\left(\phi_{l,i}\right)}
 ,\dots{},e^{j2\pi{}\left(M-1\right)d\sin\left(\phi_{l,i}\right)}\right]. 
\vspace{-2pt}
 \end{equation}
We note that $i\in\left\{1,\dots,P\right\}$, with $d$ denoting the equidistant 
inter-element spacing normalized by the carrier wavelength, $\lambda$; 
$\phi_{l,i}\in\left[-\Delta/2,\Delta/2\right]$ denotes the 
$i$-th direction-of-arrival (DOA) 
from terminal $l$ to the BS array and $\Delta$ is the angular spread in 
the azimuth domain. With such a model, the angular spread can be 
modeled by having a large $P$, whilst different degrees of receive correlation 
are adjusted by varying the angular spread. Moreover, $\bm{h}_{l}\sim{}\mathcal{CN}\left(
0,\bm{I}_{P}\right)$ is the $P\times{}1$ vector of diffuse channel gains, whilst 
$\bar{\bm{h}}_{l}$ is the $M\times{}1$ vector denoting the specular component of the 
channel and is governed by the ULA's steering response with a LoS DoA, 
$\bar{\phi}_{l}$ for terminal $l$, such that
\vspace{-3pt}
\begin{equation}
 \label{hlosterminall}
 \bar{\bm{h}}_{l}=\left[1,e^{j2\pi{}d\sin\left(\bar{\phi}_{l}\right)},\dots{},
 e^{j2\pi{}\left(M-1\right)d\sin\left(\bar{\phi_{l}}\right)}\right]. 
 \vspace{-3pt}
\end{equation}

\textbf{Remark 1.} 
For both $\bm{a}\left(\phi_{l,i}\right)$ and $\bar{\bm{h}}_{l}$, 
we note that the normalized total array length, $d_{0}$, is fixed at the BS, 
such that the inter-element spacing between two successive elements is 
given by $d=\frac{d_{0}}{M-1}\lambda$. Since the physical dimensions of the 
BS array are predetermined, the above model accurately allows us to capture the 
correlation due to close proximity of adjacent antenna elements positioned at the 
array. This along with the unique correlation matrices for each terminal created by the 
$\bm{A}_{l}$ for $l\in{}\left\{1,\dots{},L\right\}$ constitutes our focus in 
the following sections. We note that in this study, we neglect the effects of antenna 
coupling, since they can be compensated by impedance matching techniques  
as shown in \cite{WARNICK,GARCIA-ROD}. 

To determine the level of LoS and NLoS present in the propagation channel from a 
given terminal to the BS, we employ a probability based approach following 
\cite{TATARIA}. Both LoS and NLoS probabilities are a function of 
the link distance, from which the LoS and NLoS geometric attenuation, as well as 
other link characteristics are obtained. We consider propagation parameters 
from both microwave \cite{3GPPTR36873} and mmWave \cite{AKDNENIZ} frequency bands. 
For notational clarity, we delay the discussion of the above mentioned parameters to 
Section~\ref{NumericalResults}. 

\vspace{-3pt}
\subsection{Per-Terminal SINR and Ergodic Sum Spectral Efficiency}
\vspace{-2pt}
As linear signal processing techniques perform near optimally for large 
MU-MIMO systems \cite{RUSEK,LARSSON}, we employ a linear receiver in the form 
of a MRC at the BS. The $L\times{}M$ MRC matrix, $\bm{G}^{\textrm{H}}$, is used to 
separate $\bm{y}$ into $L$ streams by 
\begin{equation}
 \label{receivedsignalafterMRC}
 \bm{r}=\bm{G}^{\textrm{H}}\bm{y}=\rho^{\frac{1}{2}}\bm{G}^{\textrm{H}}\bm{G}
 \bm{D}^{\frac{1}{2}}\bm{s}+\bm{G}^{\textrm{H}}\bm{n}. 
\end{equation}
Thus, the detected signal from terminal $l$ is given by
\begin{equation}
 \label{receivedsignalterminall}
 r_{l}=\rho^{\frac{1}{2}}\beta_{l}^{\frac{1}{2}}
 \bm{g}_{l}^{\textrm{H}}\bm{g}_{l}s_{l}\hspace{-1pt}+\hspace{-1pt}
 \rho^{\frac{1}{2}}\sum\nolimits_{\substack{k=1\\k\neq{}l}}^{L}
 \hspace{-1pt}\beta_{k}^{\frac{1}{2}}\bm{g}_{l}^{\textrm{H}}
 \bm{g}_{k}s_{k}\hspace{-1pt}+\hspace{-1pt}
 \bm{g}_{l}^{\textrm{H}}\bm{n}, 
 \end{equation}
 resulting in the corresponding SINR given by 
\begin{equation}
 \label{SINRterminall}
 \textrm{SINR}_{l}=\frac{\rho\beta_{l}||\bm{g}_{l}||^{4}}{||\bm{g}_{l}||^{2}+
 \rho\sum\nolimits_{\substack{k=1\\k\neq{}l}}^{L}\beta_{k}|\bm{g}_{l}^{\textrm{H}}
 \bm{g}_{k}|^{2}}. 
\end{equation}
Hence, the instantaneous achievable uplink spectral efficiency for 
terminal $l$ (measured in bits/sec/Hz) can be computed as $\textrm{R}_{l}=
\log_{2}\left(1+\textrm{SINR}_{l}\right)$. As such, 
the ergodic sum spectral efficiency over all $L$ 
terminals is given by 
\begin{equation}
 \label{ergodicsumse}
 \mathbb{E}\left[\textrm{R}_{\textrm{sum}}\right]=\mathbb{E}\left[
 \sum\nolimits_{l=1}^{L}\hspace{-1pt}\textrm{R}_{l}\right], 
\end{equation}
where the expectation is performed over the fast-fading. In 
the following section, we derive tight analytical expressions to approximate 
the expected value of \eqref{SINRterminall} and \eqref{ergodicsumse}, respectively.

\vspace{3pt}
\section{Expected Per-Terminal SINR and Ergodic Sum Spectral Efficiency Analysis}
\label{ExpectedPerTerminalSINRandErgodicSumSpectralEff}
\vspace{3pt}
The expected SINR for terminal $l$ can be obtained by taking the expectation of 
the ratio in \eqref{SINRterminall}. However, exact evaluation of this is 
extremely cumbersome \cite{ZHANG2,BASNAYAKA}. Hence, we resort to the commonly 
used first-order Delta expansion, as shown in \cite{ZHANG2,BASNAYAKA} and 
references therein. This gives  
\begin{equation}
 \label{expectedSINRterminall}
 \mathbb{E}\left[\textrm{SINR}_{l}\right]\approx
 \frac{\rho\beta_{l}\mathbb{E}\left[||\bm{g}_{l}||^{4}\right]}
 {\mathbb{E}\left[||\bm{g}_{l}||^{2}\right]+
 \rho\sum\nolimits_{\substack{k=1\\k\neq{}l}}^{L}\beta_{k}
 \mathbb{E}\left[|\bm{g}_{l}^{\textrm{H}}\bm{g}_{k}|^{2}\right]}. 
\end{equation}

\textbf{Remark 2.} The approximation in \eqref{expectedSINRterminall} is of the 
form of $\mathbb{E}\left[\frac{X}{Y}\right]\approx\frac{\mathbb{E}\left[X\right]}
{\mathbb{E}\left[Y\right]}$. The accuracy of such approximations relies on $Y$ 
having a small variance relative to its mean. This can be seen by applying a 
multivariate Taylor series expansion of $\frac{X}{Y}$ around 
$\frac{\mathbb{E}\left[X\right]}{\mathbb{E}\left[Y\right]}$, as shown in the 
analysis methodology of \cite{ZHANG2}. In particular, both $X$ and $Y$ are 
well suited to this approximation as $M$ and $L$ start to increase (the case for 
large MU-MIMO systems), where the approximation is shown to be extremely 
tight. This is due to $X$ and $Y$ averaging their respective individual 
components, minimizing their variance relative to their mean. For further 
discussion, we refer the interested reader to Appendix I of \cite{ZHANG2}, 
where a detailed mathematical proof of the approximation accuracy can be found. 

In the sequel, Lemmas 1, 2 and 3 derive the expectations in the numerator and 
denominator of \eqref{expectedSINRterminall}. 

\textbf{Lemma 1.} For a ULA with $M$ antennas in a fixed physical 
space at the BS, considering a correlated Ricean fading channel in 
$\bm{g}_{l}$ from terminal $l$ to the BS
\vspace{-4pt}
\begin{equation}
\nonumber
\hspace{-38pt}
\delta_{l}=
\mathbb{E}\left[||\bm{g}_{l}||^{4}\right]={\left(\eta'_{l}\right)}^{4}\left\{P^{2}
M^{2}+\textrm{tr}\left[\left(\bm{A}_{l}^{\textrm{H}}
\bm{A}_{l}\right)^{2}\right]\right\}+
\vspace{-4pt}
\end{equation}
\begin{equation}
\label{lemma1}
\hspace{-1pt}
2PM^{2}\hspace{-2pt}\left(\eta'_{l}\right)^{2}\hspace{-2pt}
\left(\bar{\eta}_{l}\right)^{2}
+\hspace{2pt}2\left(\eta'_{l}\right)^{2}\hspace{-2pt}\left(\bar{\eta}_{l}\right)^{2}
\bm{\bar{h}}_{l}^{\textrm{H}}\bm{A}_{l}\bm{A}_{l}^{\textrm{H}}\bm{\bar{h}}_{l}
+\left(\bar{\eta}_{l}\right)^{4}\hspace{-2pt}M^{2}, 
\vspace{-4pt}
\end{equation}
where each parameter is defined after \eqref{terminalgchannel}. 

\emph{Proof:} See Appendix~\ref{ProofLemma1}. \QED

\textbf{Lemma 2.} Under the same conditions as Lemma 1, 
\vspace{-4pt}
\begin{equation}
\nonumber
\varphi_{l,k}=\mathbb{E}\left[|\bm{g}_{l}^{\textrm{H}}\bm{g}_{k}|^{2}\right]=
\left(\eta'_{l}\right)^{2}\hspace{-2pt}
\left(\eta'_{k}\right)^{2}\textrm{tr}\left[\hspace{-1pt}\bm{A}_{k}
\hspace{-1pt}\bm{A}_{k}^{\textrm{H}}\hspace{-1pt}
\bm{A}_{l}\hspace{-1pt}\bm{A}_{l}^{\textrm{H}}\right]+
\left(\eta'_{l}\right)^{2}\hspace{-2pt}\left(\bar{\eta}_{k}\right)^{2}
\vspace{-2pt}
\end{equation}
\begin{equation}
\nonumber
\hspace{-16pt}
\textrm{tr}\left[\bar{\bm{h}}_{k}^{\textrm{H}}\hspace{-1pt}
\bm{A}_{l}\hspace{-1pt}\bm{A}_{l}^{\textrm{H}}\hspace{-1pt}
\bar{\bm{h}}_{k}\right]\hspace{-2pt}+\hspace{-1pt}
\left(\bar{\eta}_{l}\right)^{2}\left(\eta'_{k}\right)^{2}
\hspace{-2pt}\textrm{tr}\left[\bar{\bm{h}}_{l}^{\textrm{H}}
\bm{A}_{k}\hspace{-1pt}\bm{A}_{k}^{\textrm{H}}
\hspace{-1pt}\bar{\bm{h}}_{l}\right]\hspace{-1pt}+\hspace{-1pt}
\left(\bar{\eta}_{l}\right)^{\hspace{-1pt}2}
\left(\bar{\eta}_{k}\right)^{\hspace{-1pt}2}
\end{equation}
\begin{equation}
\hspace{-200pt}
 \label{lemma2}
 \bar{\bm{h}}_{l}^{\textrm{H}}
 \bar{\bm{h}}_{k}\bar{\bm{h}}_{k}^{\textrm{H}}
 \bar{\bm{h}}_{l}. 
 \vspace{-4pt}
\end{equation}

\emph{Proof:} See Appendix~\ref{ProofLemma2}. \QED

\textbf{Lemma 3.} Under the same conditions as Lemma 1, 
\vspace{-4pt}
\begin{equation}
\label{lemma3}
\chi_{l}=\mathbb{E}\left[||\bm{g}_{l}||^{2}\right]=M\left[P\left(\eta'_{l}\right)^{2}
+\left(\bar{\eta}_{l}\right)^{2}\right]. 
\vspace{-4pt}
\end{equation}

\emph{Proof:} We begin by substituting the definition of $\bm{g}_{l}$ 
into $\chi_{l}$ and expanding, allowing us to write 
\vspace{-2pt}
\begin{equation}
\label{lemma3proof31}
\chi_{l}\hspace{-2pt}=\hspace{-2pt}\mathbb{E}\left[||\bm{g}_{l}||^{2}\right]
\hspace{-3pt}=\hspace{-2pt}
\mathbb{E}\left[\left(\eta'_{l}\right)^{2}\hspace{-2pt}\bm{h}_{l}^{\textrm{H}}
\bm{A}_{l}^{\textrm{H}}\bm{A}_{l}\bm{h}_{l}\right]\hspace{-1pt}+
\hspace{-1pt}\mathbb{E}\left[\left(\bar{\eta}_{l}\right)^{2}\hspace{-1pt}
\bar{\bm{h}}_{l}^{\textrm{H}}\bar{\bm{h}}_{l}\right]. 
\vspace{-3pt}
\end{equation}
Performing the expectations with respect to $\bm{h}_{l}$ and 
extracting the relevant constants yields 
\vspace{-2pt}
\begin{equation}
\label{lemma3proof32}
\chi_{l}=\mathbb{E}\left[||\bm{g}_{l}||^{2}\right]\hspace{-2pt}=
\hspace{-2pt}\left(\eta'_{l}\right)^{2}\textrm{tr}\left[
\bm{A}_{l}^{\textrm{H}}
\bm{A}_{l}\right]\hspace{-2pt}+\hspace{-2pt}
\left(\bar{\eta}_{l}\right)^{2}
\mathbb{E}\left[\bar{\bm{h}}_{l}^{\textrm{H}}\bar{\bm{h}}_{l}\right]. 
\vspace{-1pt}
\end{equation}
Recognizing that $\textrm{tr}\left[\bm{A}_{l}^{\textrm{H}}\bm{A}_{l}
\right]=PM$ and $\mathbb{E}\left[\bar{\bm{h}}_{l}^{\textrm{H}}
\bar{\bm{h}}_{l}\right]=M$ allows us to state
\vspace{-5pt}
\begin{equation}
\label{lemma3proof33}
\chi_{l}=\mathbb{E}\left[||\bm{g}_{l}||^{2}\right]=
M\left[P\left(\eta'_{l}\right)^{2}+\left(\bar{\eta}_{l}\right)^{2}\right], 
\vspace{-5pt}
\end{equation}
concluding the proof. \QED

\textbf{Theorem 1.} With MRC at the BS consisting of a 
space-constrained ULA, the expected uplink SINR of terminal $l$ 
in a spatially correlated Ricean fading channel can be 
approximated as 
\vspace{-3pt}
\begin{equation}
\label{theorem1}
\mathbb{E}\left[\textrm{SINR}_{l}\right]\approx
\frac{\rho\beta_{l}\delta_{l}}{\chi_{l}+\rho{}
\sum\nolimits_{\substack{k=1\\k\neq{}l}}^{L}
\beta_{k}\hspace{1pt}\varphi_{l,k}}. 
\vspace{-2pt}
\end{equation}

\emph{Proof:} Substituting the results from Lemmas 1, 2 and 3 
for $\delta_{l},\varphi_{l,k}$ and $\chi_{l}$ yields the desired 
expression in \eqref{theorem1}. \QED

\textbf{Remark 3.} The result in \eqref{theorem1} is extremely general and 
is a closed-form solution to a complex scenario, where in addition to fixed 
physical spacing and MRC at the BS, each terminal has a unique LoS direction, 
unique Rice factor, unique receive correlation matrix and a 
unique link gain. It can be readily observed via inspection, 
that both the numerator and the 
denominator of \eqref{theorem1} are influenced by each of the above factors. 
The result allows for a general evaluation of large MU-MIMO systems
with space-constrained ULAs and lends itself to many useful 
special cases (as shown in Section~\ref{SpecialCases}). 

We note that \eqref{theorem1} can be further used to approximate 
the ergodic sum spectral efficiency of the system by 
\begin{equation}
 \label{approxergse}
 \mathbb{E}\left[\textrm{R}_\textrm{sum}\right]\approx
\sum\nolimits_{l=1}^{L}\log_{2}\left(1+\mathbb{E}\left[
\textrm{SINR}_{l}\right]\right).
\end{equation}
The accuracy of the derived closed-form expressions in \eqref{theorem1} and 
\eqref{approxergse} is demonstrated in Section~\ref{NumericalResults}. 
In the following section, we present three special cases of Theorem 1 
demonstrating its generality.

\vspace{-5pt}
\section{Special Cases}
\label{SpecialCases}
\textbf{Corollary 1.} With MRC at the BS consisting of a SC ULA, 
the expected uplink SINR of terminal $l$ with 
no LoS, i.e, Rayleigh fading with unequal correlation matrices for 
each terminal, can be approximated as 
\vspace{4pt}
\begin{equation}
 \nonumber
 \hspace{-190pt}
\mathbb{E}\left[\textrm{SINR}_{l}^{\textrm{C1}}\right]
\hspace{-3pt}\approx\hspace{-2pt}
\vspace{-7pt}
\end{equation}
\begin{equation}
 \label{corollary1}
 \hspace{8pt}
 \frac{\rho\beta_{l}\left(\eta'_{l}\right)^{4}\left\{
 P^{2}M^{2}\hspace{-1pt}+\hspace{-1pt}
 \textrm{tr}\left[\left(\bm{A}_{l}^{\textrm{H}}\bm{A}_{l}
 \right)^{2}\right]\right\}}
 {M\hspace{-2pt}P\hspace{-2pt}\left(\eta'_{l}\right)^{2}\hspace{-2pt}+
 \hspace{-2pt}\rho
 \sum\nolimits_{\substack{k=1\\k\neq{}l}}^{L}
 \left\{\beta_{k}\left(\eta'_{l}
 \right)^{2}\hspace{-1pt}\left(\eta'_{k}\right)^{2}
 \hspace{-1pt}\textrm{tr}\left[
 \hspace{-1pt}\bm{A}_{k}\hspace{-1pt}\bm{A}_{k}^{\textrm{H}}
 \hspace{-1pt}\bm{A}_{l}\hspace{-1pt}\bm{A}_{l}^{\textrm{H}}
 \right]\right\}}.
 \vspace{-2pt}
\end{equation}

\emph{Proof:} Substituting $\delta_{l}$, $\chi_{l}$ and 
$\varphi_{l}$ into \eqref{theorem1} and setting 
$\bar{\eta}_{l}\hspace{-1pt}=\hspace{-1pt}
\bar{\eta}_{k}=0$, $\eta'_{l}\hspace{-1pt}=\hspace{-1pt}
\eta'_{k}=\frac{1}{\sqrt{P}}$ as $K_{l}=0$ and 
$\bar{\bm{h}}_{l}=\bar{\bm{h}}_{k}=\bm{0}_{M\times{}1}$, 
where $\bm{0}_{M\times{}1}$ denotes a $M\times{}1$ vector of 
zeros for $l,k\in\left\{1,\dots{},L\right\}$ yielding 
the desired expression. \QED

\vspace{2pt}
\textbf{Corollary 2 (Proposition 1 in \cite{ZHANG}).} 
With MRC processing at the BS containing of a SC 
ULA, the expected uplink SINR for terminal $l$ with 
no LoS and equal correlation matrices, i.e., 
Rayleigh fading with a fixed correlation for each 
terminal, can be approximated as 
\vspace{4pt}
\begin{equation}
\nonumber
\hspace{-190pt}
 \mathbb{E}\left[\textrm{SINR}_{l}^{\textrm{C2}}\right]
\hspace{-3pt}\approx\hspace{-2pt}
\vspace{-7pt}
\end{equation}
\begin{equation}
 \label{corollary2}
 \frac{\rho\beta_{l}\left(\eta'_{l}\right)^{4}\left\{
 P^{2}M^{2}\hspace{-1pt}+\hspace{-1pt}
 \textrm{tr}\left[\left(\bm{A}_{l}^{\textrm{H}}\bm{A}_{l}
 \right)^{2}\right]\right\}}
 {M\hspace{-2pt}P\hspace{-2pt}\left(\eta'_{l}\right)^{2}\hspace{-2pt}+
 \hspace{-2pt}\rho
 \sum\nolimits_{\substack{k=1\\k\neq{}l}}^{L}
 \left\{\beta_{k}\left(\eta'_{l}
 \right)^{2}\hspace{-1pt}\left(\eta'_{k}\right)^{2}
 \textrm{tr}\left[\left(\bm{A}_{l}^{\textrm{H}}\bm{A}_{l}
 \right)^{2}\right]\right\}}.
 \vspace{-3pt}
\end{equation}

\emph{Proof:} Following the approach outlined in the proof of 
Corollary 1 and recognizing that $\textrm{tr}\big[\hspace{-2pt}\left(
\hspace{-1pt}\bm{A}_{l}\hspace{-1pt}
\bm{A}_{l}^{\textrm{H}}\right)^{2}\big]=\textrm{tr}
\left[\bm{A}_{l}\hspace{-1pt}
\bm{A}_{l}^{\textrm{H}}\hspace{-1pt}
\bm{A}_{l}\hspace{-1pt}\bm{A}_{l}^{\textrm{H}}
\right]\hspace{-2pt}=\hspace{-2pt}\textrm{tr}
\big[\hspace{-1pt}\left(\hspace{-1pt}
\bm{A}_{l}^{\textrm{H}}\hspace{-1pt}\bm{A}_{l}\right)^{2}\big]$ 
yields the desired result. \QED

\vspace{2pt}
\textbf{Corollary 3.} With MRC at the BS consisting of a SC ULA, 
the expected uplink SINR of terminal $l$ with 
LoS i.e., correlated Ricean fading, with equal correlation 
matrices for each terminal can be approximated as 
\begin{equation}
\label{corollary3}
\mathbb{E}\left[\textrm{SINR}_{l}^{\textrm{C3}}\right]\approx
\frac{\rho\beta_{l}\delta_{l}}{\chi_{l}+\rho
\sum\nolimits_{\substack{k=1\\k\neq{}l}}^{L}
\beta_{k}\varphi'_{l,k}}, 
\vspace{-10pt}
\end{equation}
where
\vspace{-2pt}
\begin{equation}
\nonumber
\varphi'_{l,k}=\left(\eta'_{l}\right)^{2}\hspace{-2pt}
\left(\eta'_{k}\right)^{2}\textrm{tr}\big[\hspace{-1pt}
\left(\hspace{-2pt}\bm{A}_{l}
\hspace{-1pt}\bm{A}_{l}^{\textrm{H}}\right)^{2}\big]\hspace{-2pt}+\hspace{-2pt}
\left(\eta'_{l}\right)^{2}\hspace{-2pt}\left(\bar{\eta}_{k}\right)^{2}
\textrm{tr}\left[\bar{\bm{h}}_{k}^{\textrm{H}}\hspace{-1pt}
\bm{A}_{l}\hspace{-1pt}\bm{A}_{l}^{\textrm{H}}\hspace{-1pt}
\bar{\bm{h}}_{k}\right]
\vspace{-3pt}
\end{equation}
\begin{equation}
\label{varphiprimel}
\hspace{-52pt}
\hspace{-2pt}+\hspace{-1pt}
\left(\bar{\eta}_{l}\right)^{2}\left(\eta'_{k}\right)^{2}
\hspace{-1pt}\textrm{tr}\left[\bar{\bm{h}}_{l}^{\textrm{H}}\hspace{-2pt}
\bm{A}_{l}\hspace{-1pt}\bm{A}_{l}^{\textrm{H}}\bar{\bm{h}}_{l}\right]
\hspace{-2pt}+\hspace{-2pt}\left(\bar{\eta}_{l}\right)^{\hspace{-1pt}2}
\left(\bar{\eta}_{k}\right)^{\hspace{-1pt}2}M^{2}. 
\vspace{-5pt}
\end{equation} 

\emph{Proof:} Replacing $\textrm{tr}\left[\bm{A}_{k}\hspace{-1pt}
\bm{A}_{k}^{\textrm{H}}\hspace{-1pt}\bm{A}_{l}\hspace{-1pt}
\bm{A}_{l}^{\textrm{H}}\right]$ with  
$\textrm{tr}\big[\hspace{-2pt}
\left(\hspace{-2pt}\bm{A}_{l}\bm{A}_{l}^{\textrm{H}}\right)^{2}
\big]$ yields the desired expression in \eqref{corollary3}. 
We note that $\delta_{l}$ and $\chi_{l}$ are as 
defined in \eqref{lemma1} and \eqref{lemma3}, respectively. \QED

\textbf{Remark 4.} Corollaries 1 and 2 share 
a common trend in that both the numerators and denominators are 
governed by spatial correlation matrices in $\bm{A}_{l}$ and $\bm{A}_{k}$, 
respectively. In the case where correlation matrices are fixed for each 
terminal, the trace in their respective denominators 
can be readily seen to translate from 
$\textrm{tr}\left[\bm{A}_{k}\hspace{-1pt}\bm{A}_{k}^{\textrm{H}}
\hspace{-1pt}\bm{A}_{l}\hspace{-1pt}\bm{A}_{l}^{\textrm{H}}\right]$ to 
$\textrm{tr}\left[(\bm{A}_{l}^{\textrm{H}}\hspace{-1pt}\bm{A}_{l})^{2}\right]$. 

In the subsequent section, we analyze the convergence of the expected 
per-terminal SINR and ergodic spectral efficiency with MRC,  
as the number of receive antennas, $M$, grows without bound with a fixed 
number of user terminals, $L$.

\section{Limiting Expected Per-Terminal SINR and Ergodic Sum 
Spectral Efficiency Analysis}
\label{LimitingPerTerminalSINRandErgodicSumSpectralEff}
Theorem 1 presents an expected uplink SINR approximation for terminal $l$ 
which is suitable for any system size, as well as 
any operating SNR, LoS level, spatial correlation level and 
physical array spacing. We now examine the 
asymptotic behavior of \eqref{theorem1}, as $M\rightarrow\infty$, with a 
fixed (finite) $L$. Dividing through by $M^{2}$ throughout, we observe 
the limit as 
\begin{equation}
 \label{limitingSINR2}
 \overline{\mathbb{E}\left[\textrm{SINR}_{l}\right]}=
 \lim_{M\rightarrow{}\infty}
  \Bigg\{\frac{\rho\beta_{l}\left(\delta_{l}/M^{2}\right)}
 {\left(\chi_{l}/M^{2}\right)+
 \rho\sum\nolimits_{\substack{k=1\\k\neq{}l}}^{L}
 \beta_{k}\left(\varphi_{l,k}/M^{2}\right)}\Bigg\}. 
 \vspace{-6pt}
\end{equation}
Referring to the numerator of \eqref{limitingSINR2}, two 
terms in 
\vspace{-1pt}
\begin{equation}
 \delta_{l}^{1}=\left(\eta'_{l}\right)^{4}
(\textrm{tr}[(\bm{A}_{l}^{\textrm{H}}
\hspace{-1pt}\bm{A}_{l})^{2}]/M^{2}),
\vspace{-3pt}
\end{equation}
and 
\begin{equation}
\vspace{-3pt}
 \delta_{l}^{2}=2\left(\eta'_{l}\right)^{2}
\left(\bar{\eta}_{l}\right)^{2}
(\bar{\bm{h}}_{l}^{\textrm{H}}
\bm{A}_{l}\hspace{-1pt}\bm{A}_{l}^{\textrm{H}}
\bar{\bm{h}}_{l}/M^{2}),
\end{equation}
do not vanish from $\delta_{l}$ as $M$ grows without bound, 
whilst the denominator of \eqref{limitingSINR2} has four 
terms, these are
\vspace{-1pt}
\begin{equation}
 \varphi_{l,k}^{1}=\left(\eta'_{l}\right)^{2}
\left(\eta'_{k}\right)^{2}(\textrm{tr}\left[\bm{A}_{k}
\bm{A}_{k}^{\textrm{H}}\bm{A}_{l}
\bm{A}_{l}^{\textrm{H}}\right]\hspace{-2pt}/M^{2}), 
\vspace{-3pt}
\end{equation}
\begin{equation}
 \varphi_{l,k}^{2}=
\left(\eta'_{l}\right)^{2}
\left(\bar{\eta}_{k}\right)^{2}(\textrm{tr}\left[
\bar{\bm{h}}_{k}^{\textrm{H}}
\bm{A}_{l}\bm{A}_{l}^{\textrm{H}}\bar{\bm{h}}_{k}\right]/M^{2}), 
\vspace{-3pt}
\end{equation}
\begin{equation}
 \varphi_{l,k}^{3}=
\left(\bar{\eta}_{l}\right)^{2}\left(\eta'_{k}\right)^{2}
(\textrm{tr}\left[\bar{\bm{h}}_{l}^{\textrm{H}}\bm{A}_{k}
\hspace{-2pt}\bm{A}_{k}^{\textrm{H}}
\bar{\bm{h}}_{l}\right]/M^{2}), 
\vspace{-3pt}
\end{equation}
and 
\vspace{-3pt}
\begin{equation}
 \varphi_{l,k}^{4}=\left(\bar{\eta}_{l}\right)^{2}
\left(\bar{\eta}_{k}\right)^{2}(
\bar{\bm{h}}_{l}^{\textrm{H}}\bar{\bm{h}}_{k}
\bar{\bm{h}}_{k}^{\textrm{H}}\bar{\bm{h}}_{l}/M^{2}), 
\vspace{-4pt}
\end{equation}
which do not vanish from $\varphi_{l,k}$ as $M\rightarrow\infty$. 

In the sequel, Lemmas 4, 5 and 6 derive the limiting value of (24)-(29), 
respectively. 

\textbf{Lemma 4.} $\lim\limits_{M\rightarrow{}\infty}\varphi_{l,k}^{4}$ 
is given by 
\vspace{-8pt}
\begin{equation}
 \nonumber
 \bar{\varphi}_{l,k}^{4}=
 \left(\bar{\eta}_{l}\right)^{2}
 \left(\bar{\eta}_{k}\right)^{2}
  \lim_{M\rightarrow{}\infty}\left\{\hspace{2pt}\left|
 \frac{\bar{\bm{h}}_{l}^{\textrm{H}}\bar{\bm{h}}_{k}}
 {M}\right|^{2}\right\}
 \vspace{-3pt}
\end{equation}
\begin{equation}
 \label{lemma4}
 \hspace{-8pt}
 =\left(\bar{\eta}_{l}\right)^{2}
 \left(\bar{\eta}_{k}\right)^{2}
 \vartheta\left(\bar{\phi}_{l},
 \bar{\phi}_{k}\right)^{2},
 \vspace{-3pt}
\end{equation}
where $\vartheta\left(\bar{\phi}_{l},\bar{\phi}_{k}
\right)=\left|\textrm{sinc}\left(\pi{}d_{0}\left(
\sin\left(\bar{\phi}_{l}\right)-
\sin\left(\bar{\phi}_{k}\right)\right)\big/\lambda\right)
\right|,$ where $\textrm{sinc}\left(\cdot\right)$ denotes the 
sinc function. 

\vspace{1pt}
\emph{Proof:} We begin by defining 
\vspace{-2pt}
\begin{equation}
 \nonumber
 \hspace{-120pt}
 \vartheta\left(\bar{\phi}_{l},\bar{\phi}_{k}
 \right)=\lim_{M\rightarrow{}\infty}\left\{\left|
 \frac{\bar{\bm{h}}_{l}^{\textrm{H}}\bar{\bm{h}}_{k}}
 {M}\right|\right\}
 \vspace{-3pt}
\end{equation}
\begin{equation}
\nonumber
\hspace{40pt}
 =\lim_{M\rightarrow\infty}\left\{\left|
 \frac{1}{M}\sum\nolimits_{c=0}^{M-1}e^{j2\pi{}\frac{c}
 {\lambda}\frac{d_{0}}{M-1}\left(\sin\left(\bar{\phi_{l}}
 \right)-\sin\left(\bar{\phi_{k}}\right)\right)}\right|\right\}
\vspace{-3pt}
 \end{equation}
\begin{equation}
\nonumber
\hspace{-18pt}
 =\left|\int\nolimits_{0}^{1}e^{j2\pi{}\frac{d_{0}}
 {\lambda}\left(\sin\left(\bar{\phi}_{l}\right)
 -\sin\left(\bar{\phi}_{k}\right)\right)f}df\right|
 \vspace{-3pt}
 \end{equation}
\begin{equation}
 \label{lemma4proof1}
 \hspace{5pt}
 =|\textrm{sinc}\left(\pi{}d_{0}\left(\sin\left(
 \bar{\phi_{l}}\right)-\sin\left(\bar{\phi}_{k}
 \right)\right)/\lambda\right)|, 
 \vspace{-5pt}
\end{equation}
yielding the desired result. \QED

\textbf{Remark 5.} The expression in \eqref{lemma4} is another 
closed-form solution and can be readily seen to be 
dependent on the respective LoS angles unique to terminals $l$ and $k$.

\textbf{Lemma 5.} $\lim\limits_{M\rightarrow\infty}\varphi_{l,k}^{3}$ is 
given by 
\vspace{-5pt}
\begin{equation}
 \nonumber
 \bar{\varphi}_{l,k}^{3}=
 \left(\bar{\eta}_{l}\right)^{2}
 \left(\eta'_{k}\right)^{2}
 \lim_{M\rightarrow{}\infty}
 \left\{\frac{\textrm{tr}
 \left[\bar{\bm{h}}_{l}^{\textrm{H}}
 \bm{A}_{k}\bm{A}_{k}^{\textrm{H}}\bar{\bm{h}}_{l}\right]}{M^{2}}
 \right\}
 \vspace{-8pt}
\end{equation}
\begin{equation}
 \label{lemma5}
 \hspace{-16pt}
 =\left(\bar{\eta}_{l}\right)^{2}\hspace{-2pt}
 \left(\eta'_{k}\right)^{2}
 \sum\limits_{r=1}^{P}
 \vartheta\left(\bar{\phi}_{l},\phi_{k,r}\right)^{2}.
 \vspace{-4pt}
\end{equation}

\emph{Proof:} Using similar methodology as in the proof 
of Lemma 4, we recognize that 
\vspace{-5pt}
\begin{equation}
 \label{lemma5proof1}
 \frac{1}{M^{2}}\hspace{2pt}\textrm{tr}\left[\bm{h}_{l}^{\textrm{H}}\bm{A}_{k}
 \bm{A}_{k}^{\textrm{H}}\bm{h}_{l}\right]
 \hspace{-2pt}=\hspace{-2pt}
 \frac{1}{M^{2}}\sum\limits_{r=1}^{P}\left|
 \bar{\bm{h}}_{l}^{\textrm{H}}\bm{A}_{k}\right|^{2}. 
 \vspace{-5pt}
\end{equation}
Substituting the specular and diffuse angles, $\bar{\phi}_{l}$ and 
$\phi_{k,r}$ with $r\in\left\{1,\dots{},P\right\}$, corresponding to 
$\bar{\bm{h}}_{l}$ and $\bm{A}_{k}$, yields \eqref{lemma5proof1}. \QED

\textbf{Remark 6.} We note that as $\varphi_{l,k}^{2}$ and 
$\delta_{l}^{2}$ have a similar 
structure to $\varphi_{l,k}^{3}$, 
the limiting values of $\varphi_{l,k}^{2}$ and $\delta_{l}^{2}$ in 
$\bar{\varphi}_{l,k}^{2}$ and $\bar{\delta}_{l}^{2}$ have the 
same form as \eqref{lemma5}, except the angles in 
$\vartheta\left(\cdot\right)$ are replaced with $\bar{\phi}_{k}$, 
$\phi_{l,r}$ for $\bar{\varphi}_{l,k}^{2}$ and $\bar{\phi}_{l}$, $\phi_{l,r}$ 
for $\bar{\delta}_{l}^{2}$, respectively. We further note that both 
$\bar{\varphi}_{l,k}^{2}$ and $\bar{\delta}_{l}^{2}$ will need 
to have the necessary scaling of $\left(\eta'_{l}\right)^{2}\left(\bar{\eta}_{k}
\right)^{2}$ and $2\left(\eta'_{l}\right)^{2}\left(\bar{\eta_{l}}
\right)^{2}$ as shown in (27) and (25). 

\textbf{Lemma 6.} $\lim\limits_{M\rightarrow{}\infty}\varphi_{l,k}^{1}$ 
is given by 
\vspace{-6pt}
\begin{equation}
 \nonumber
 \hspace{-17pt}
 \bar{\varphi}_{l,k}^{1}=\left(\eta'_{l}\right)^{2}
 \left(\eta'_{k}\right)^{2}\lim_{M\rightarrow{}\infty}
 \left\{\frac{\textrm{tr}\left[\bm{A}_{k}\bm{A}_{k}^{\textrm{H}}
 \bm{A}_{l}\bm{A}_{l}^{\textrm{H}}\right]}{M^{2}}\right\}
 \vspace{-6pt}
\end{equation}
\begin{equation}
 \label{lemma6}
 \hspace{-20pt}
  =\left(\eta'_{l}\right)^{2}
 \left(\eta'_{k}\right)^{2}\sum\limits_{r=1}^{P}
 \sum\limits_{t=1}^{P}\vartheta\left(\phi_{k,r},
 \phi_{l,t}\right)^{2}.
\end{equation}

\vspace{-2pt}
\emph{Proof:} Manipulating the trace in \eqref{lemma6} 
allows us to state 
\vspace{-2pt}
\begin{equation}
 \nonumber
 \frac{1}{M^{2}}\left\{\textrm{tr}\left[\bm{A}_{k}\bm{A}_{k}^{\textrm{H}}
 \bm{A}_{l}\bm{A}_{l}^{\textrm{H}}\right]\right\}=
 \frac{1}{M^{2}}\left\{\textrm{tr}\left[\bm{A}_{k}^{\textrm{H}}
 \bm{A}_{l}\bm{A}_{l}^{\textrm{H}}\bm{A}_{k}\right]\right\}
 \vspace{-3pt}
\end{equation}
\begin{equation}
 \label{prooflemma61}
 \hspace{-60pt}
 \hspace{-1pt}=\hspace{-1pt}\frac{1}{M^{2}}
 \sum\limits_{r=1}^{P}\sum\limits_{t=1}^{P}
 \left|\bm{a}^{\textrm{H}}\left(\phi_{k,r}\right)
 \bm{a}\left(\phi_{l,t}\right)\right|^{2}\hspace{-3pt}.  
 \vspace{-3pt}
\end{equation}
Substituting the respective angles and performing some routine 
algebra yields the desired result. \QED

\textbf{Remark 7.} We note that as $\delta_{l}^{1}$ has a similar 
form to $\varphi_{l,k}^{1}$. Using the same methodology as in 
Lemma 6, we can obtain $\bar{\delta}_{l}^{1}$, the limiting value 
of $\delta_{1}^{1}$, where the angles in 
$\vartheta\left(\cdot\right)$ are replaced by $\phi_{l,r},\phi_{l,t}$ 
with $\left(\eta'_{l}\right)^{4}$ providing the required scaling. 

\textbf{Theorem 2.} The limiting uplink SINR for terminal $l$ 
with MRC and a SC ULA at the BS can be written as 
\begin{equation}
 \label{theorem2}
 \overline{\mathbb{E}\left[\textrm{SINR}_{l}\right]}=
 \frac{\rho\beta_{l}\left(\bar{\delta}_{l}^{1}\hspace{1pt}+
 \bar{\delta}_{l}^{2}\right)}
 {\rho\sum\limits_{\substack{k=1\\k\neq{}l}}^{L}\beta_{k}
 (\bar{\varphi}_{l,k}^{1}+\bar{\varphi}_{l,k}^{2}+\bar{\varphi}_{l,k}^{3}
 +\bar{\varphi}_{l,k}^{4})}. 
\vspace{-6pt}
\end{equation}

\emph{Proof:} Using the results from Lemmas 4, 5, 6 and keeping in mind 
Remarks 6 and 7 yields the desired expression. \QED

As such the limiting ergodic sum spectral efficiency is given by 
\vspace{-7pt}
\begin{equation}
 \label{limitingergse}
 \overline{\mathbb{E}\left[\textrm{R}_{\textrm{sum}}\right]}=
 \sum\limits_{l=1}^{L}\log_{2}\left(1+
 \overline{\mathbb{E}\left[\textrm{SINR}_{l}\right]}\right). 
 \vspace{-5pt}
\end{equation}

In the following section, we demonstrate the accuracy of the 
analysis presented in Sections~\ref{ExpectedPerTerminalSINRandErgodicSumSpectralEff}, 
\ref{SpecialCases} and 
\ref{LimitingPerTerminalSINRandErgodicSumSpectralEff}, respectively.

\vspace{-4pt}
\section{Numerical Results}
\label{NumericalResults}
Unless otherwise specified, the parameters used in the numerical 
results are specified in Table~\ref{Table1} for an UMi scenario. 
The parameters for microwave and mmWave frequencies were obtained 
from from \cite{3GPPTR36873} and \cite{AKDNENIZ}, respectively. 
A circular cell of radius $100$ m is considered with an exclusion 
radius of $r_{0}=10$ m. We assume a uniform distribution of terminals in 
the cell area and consider $10^{4}$ Monte-Carlo realizations for 
each result. The parameter $\varrho$ is chosen such that the fifth 
percentile value of the instantaneous per-terminal SINR is 0 dB 
at SNR ($\rho$) = 0 dB for the system dimensions of 
$M=256$ and $L=32$. 
\vspace{-3pt}
\begin{table}[!h]
\begin{center}
\scalebox{0.8}{\begin{tabular}{ccc}
\toprule
\toprule
{\textbf{Parameter}} & \multicolumn{2}{c}{\textbf{Value}} \\
\cmidrule{2-3}
{} & {\textbf{Microwave}} & {\textbf{mmWave}} \\
\midrule
Carrier frequency [GHz] & $2$ & $28$\\ 
LoS attenuation exponent $\left[\alpha\right]$ & 2.2  & 2  \\ 
NLoS attenuation exponent &3.67 &2.92 \\ 
LoS shadow fading standard deviation $\left[\sigma_\textrm{sh}\right]$ 
&3 &5.8 \\ 
NLoS shadow fading standard deviation & 4 & 8.7 \\ 
$K$-Factor mean [dB] & 9 & 12 \cite{THOMAS}\\
$K$-Factor standard deviation [dB] & 5 & 3 \cite{THOMAS} \\
\bottomrule
\end{tabular}}
\end{center}
\caption{System Parameters}
\label{Table1}
\vspace{-21pt}
\end{table}

Based on the link distance, $r_{l}$, we employ a probability based approach 
in determining whether the terminal experiences LoS or NLoS conditions on 
the uplink to the BS. For the microwave case, the probability of terminal $l$ 
experiencing LoS is governed by $P_{\textrm{LoS}}\left(r_{l}\right)=
(\min\left(18/r_{l},1\right)\big(1-e^{-r_{l}/36}
\big))+e^{-r_{l}/36}$. Naturally, the probability of the terminal 
experiencing NLoS is then determined 
by $P_{\textrm{NLoS}}=1-P_{\textrm{LoS}}$. 
Equivalently, for the mmWave case \cite{AKDNENIZ}, 
$P_{\textrm{LoS}}\left(r_{l}\right)=\left(1-P_{\textrm{out}}
 \left(r_{l}\right)\right)e^{-\omega{}_{\textrm{LoS}}\hspace{2pt}r_{l}}$,
where $1/\omega_{\textrm{LoS}}=67.1$ meters and $P_{\textrm{out}}$ is the 
outage probability, occurring when the attenuation in either the LoS 
or NLoS states is sufficiently large. For simplicity, we set 
$P_{\textrm{out}}=0$ when determining the LoS and NLoS probabilities. 
Upon determining the link state of each terminal, we select the corresponding 
link parameters to model the large-scale propagation effects of geometric 
attenuation and shadow-fading, as specified in Table~
\ref{Table1}. We assign a 
unique $K$-factor, $K_{l}$, for the $l$-th user terminal from a log-normal 
distribution with the mean and standard deviation specified in 
Table~\ref{Table1}. We refer to this as $K_{l}
\sim\textrm{ln}\left(\textrm{mean},\textrm{standard deviation}\right)$.

\begin{figure}[!t]
 \centering
 \vspace{-15pt}
 \includegraphics[width=8.3cm]{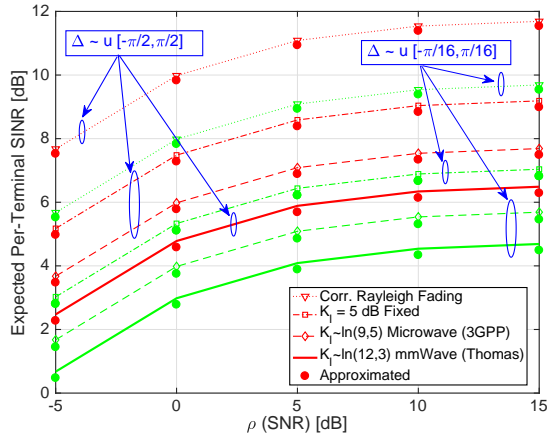}
 \vspace{-12pt}
 \caption{Expected per-terminal SINR vs. $\rho$ (SNR) 
 with $M=256,L=32,P=50,d_{0}=8\lambda$.} 
 \label{expsinrvssnrF1}
\end{figure}

First, the accuracy of the proposed expected per-terminal SINR in 
\eqref{theorem1} is examined. Fig.~\ref{expsinrvssnrF1} illustrates 
the expected SINR of a given terminal as a function of $\rho$ (SNR) 
for a system with $M=256$ and $L=32$, $P=50$ and $d_{0}=8\lambda$. 
In addition to the microwave and 
mmWave cases, we consider the correlated Rayleigh fading case  
for comparison purposes. We also consider the case where each terminal 
is assigned a fixed $K$-factor of $5$ dB. Three trends can be observed: 
Firstly, transitioning from large to small angular spread 
($\Delta\sim{}u[\frac{-\pi}{2},\frac{\pi}{2}]$ 
to $\Delta\sim{}u[\frac{-\pi}{16},\frac{\pi}{16}])$ 
tends to significantly reduce the expected SINR for all cases. This is 
despite the fact that the ULA contains 
very large numbers of antenna elements at the BS, 
and is due to the reduction in the spatial diversity (rank) of the channel,   
allowing the BS array to only see a very narrow spread of 
incoming power. Secondly, increasing the mean of $K$ has an adverse effect 
on the expected SINR. This is because a stronger specular component in 
the channel tends to reduce the multipath diversity and in-turn reduces 
its overall rank. Equivalently, this can be interpreted as an increase in 
the level of inter-terminal interference leading to a lower expected 
per-terminal SINR. 
Third, our proposed approximations are seen to remain extremely tight 
for the entire SNR range for all cases. The analytical expressions are also 
seen to 
remain tight for the special case presented in \eqref{corollary1}, where 
each terminal undergoes Rayleigh fading with unequal correlation matrices. 
Furthermore, the expected SINR in each case is seen to saturate with 
growing SNR, due to the inability of the MRC to mitigate inter-terminal 
interference. 

\begin{figure}[!t]
 \centering
 \vspace{-14pt}
 \includegraphics[width=8.3cm]{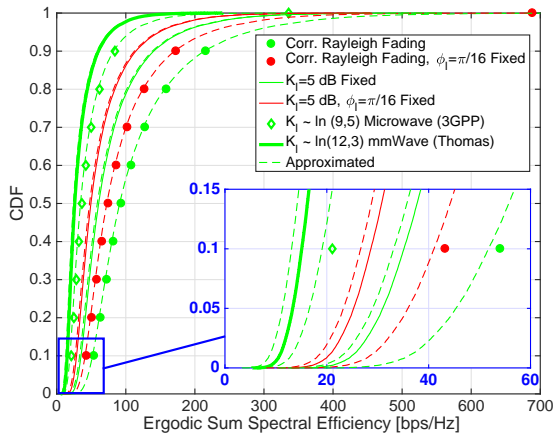}
 \vspace{-8pt}
 \caption{Ergodic sum spectral efficiency CDF with $M=256, L=32$, $\rho
 \hspace{3pt}(\textrm{SNR})=10$ dB, $P=50, d_{0}=8\lambda$ and 
 $\Delta\sim{}u[-\frac{\pi}{16},\frac{\pi}{16}]$ (unless specified in 
 the figure).} 
 \label{ergseCDFF2}
\end{figure}
\begin{figure}[!t]
 \centering
 \vspace{-10pt}
 \includegraphics[width=8.3cm]{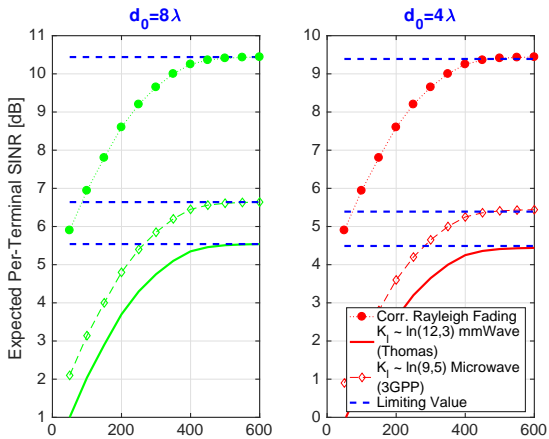}
 \vspace{-12pt}
 \caption{Expected Per-Terminal SINR vs. $M$ with $L=32$ at 
 $\rho$ (SNR) $=10$ dB, $P=50$, 
 $\Delta\sim{}u[-\frac{\pi}{16},\frac{\pi}{16}]$.} 
 \label{limitsF3}
 \vspace{-20pt}
\end{figure}
Considering the special cases in \eqref{corollary2} and \eqref{corollary3}, 
we now examine the influence of LoS, as well as equal and unequal 
correlation matrices on the ergodic sum spectral efficiency, as 
shown in Fig.~\ref{ergseCDFF2}. Using the same propagation parameters from 
Fig.~\ref{expsinrvssnrF1}, 
(listed in the figure caption) at $\rho$ (SNR) $\hspace{-3pt}=10$ dB, 
we compare the cumulative distribution functions (CDFs) 
of the derived ergodic sum spectral efficiency approximation in 
\eqref{approxergse} with its simulated counterparts. We note that the 
CDF is obtained by averaging over the 
fast-fading in the channel with each value representing the variations 
in the link gains and $K$-factors. We notice that 
irrespective of the underlaying propagation characteristics 
(Rayleigh or Ricean fading), unequal correlation matrices results in a 
higher ergodic sum spectral efficiency of the system allowing the ULA to 
leverage a larger amount of spatial diversity. Furthermore, we again 
observe that a stronger specular component tends to decrease the ergodic 
sum spectral efficiency. The derived approximations are robust to the 
presence of equal and unequal correlation matrices, as well as changes in 
the level of LoS. 
We also evaluate the accuracy of the limiting expected SINR expression 
derived in \eqref{theorem2}, with growing numbers of BS antennas and a 
fixed number of terminals in the system at $L=32$. Three trends can be 
observed: After recognizing that increasing $M$ increases the expected 
SINR, for each case the expected SINR slowly 
saturates with growing $M$ and approaches its limiting value at 
approximately 500 antenna elements for each case, respectively. This is a 
result of channels from multiple terminals becoming asymptotically 
orthogonal. Secondly, decreasing the physical size of the array further 
reduces the inter-element spacing translating into a reduction in the 
expected SINR for all cases respectively. Finally, we can observe 
that each case converges to the derived limiting value.

\vspace{-5pt}
\section{Conclusion}
\label{Conclusion}
\vspace{-3pt}
In this paper, we investigated the uplink performance of large MU-MIMO 
systems under spatially correlated Ricean fading, 
with ULAs at the BS employed in a fixed physical space. Closed-form 
approximations to the expected per-terminal SINR and ergodic sum spectral 
efficiency are derived with MRC processing at the BS. In the limit of a 
large number of BS antennas, asymptotic expressions for the expected 
per-terminal SINR 
and ergodic sum spectral efficiency were derived. Our numerical results 
show that with constraints on the physical size of the ULA, the expected 
SINR saturated with increasing SNR and BS antenna numbers. The analysis 
accommodates to changes in system dimensions, operating SNR, LoS levels, 
spatial correlation levels and variation in fixed physical spacings. 
Unequal correlation matrices to each terminal resulted in a performance 
increase, whilst LoS had an adverse impact on system performance.

\vspace{-5pt}
\appendices
\section{Proof of Lemma 1}
\label{ProofLemma1}
\vspace{-2pt}
We begin by recognizing 
$\delta_{l}=\mathbb{E}\left[||\bm{g}_{l}||^{4}\right]=
\mathbb{E}\big[\hspace{-2pt}
\left(||\bm{g}_{l}||^{2}\right)^{2}\big]$. 
Substituting the definition of $\bm{g}_{l}$ and 
denoting $\bm{v}_{l}=\eta'_{l}\bm{A}_{l}\bm{h}_{l}$ and 
$\bm{q}_{l}=\bar{\eta}_{l}\bar{\bm{h}}_{l}$ allows us to 
state 
\vspace{-3pt}
\begin{equation}
 \label{prooflemma11}
 \delta_{l}\hspace{-2pt}=\hspace{-2pt}
\mathbb{E}\big[\hspace{-2pt}\left(||\bm{g}_{l}||^{2}\right)^{2}\big]
\hspace{-2pt}=\hspace{-2pt}\mathbb{E}\big[\hspace{-2pt}
 \left(\bm{v}_{l}^{\textrm{H}}\bm{v}_{l}\hspace{-2pt}+\hspace{-2pt}
 \bm{v}_{l}^{\textrm{H}}\bm{q}_{l}\hspace{-2pt}+\hspace{-2pt}
 \bm{q}_{l}^{\textrm{H}}\bm{v}_{l}\hspace{-2pt}+\hspace{-2pt}
 \bm{q}_{l}^{\textrm{H}}\bm{q}_{l}
 \right)^{2}\big].
 \vspace{-6pt}
\end{equation}
Expanding \eqref{prooflemma11} and simplifying allows us to state 
\vspace{-4pt}
\begin{equation}
 \nonumber
 \hspace{-4pt}
  \delta_{l}=\mathbb{E}\left[\left(||\bm{g}_{l}||^{2}\right)^{2}\right]=
  \mathbb{E}\big[\hspace{-2pt}\left(\bm{v}_{l}^{\textrm{H}}
  \bm{v}_{l}\right)^{2}\big]
  +\mathbb{E}\big[2\left(\bm{v}_{l}^{\textrm{H}}\bm{v}_{l}\right)
  \left(\bm{q}_{l}^{\textrm{H}}\bm{q}_{l}\right)\big]+
  \vspace{-6pt}
\end{equation}
\begin{equation}
  \label{proof1lemma12}
  \hspace{-63pt}
  \mathbb{E}\left[\bm{v}_{l}^{\textrm{H}}\bm{q}_{l}\bm{q}_{l}^{\textrm{H}}
  \bm{v}_{l}\right]+\mathbb{E}\big[\bm{q}_{l}^{\textrm{H}}\bm{v}_{l}
  \bm{v}_{l}^{\textrm{H}}\bm{q}_{l}\big]+
  \mathbb{E}\big[\big(\bm{q}_{l}^{\textrm{H}}\bm{q}_{l}\big)^{2}\big]. 
  \vspace{-3pt}
\end{equation}
Performing the expectation over $\bm{v}_{l}$ in the last four
terms of \eqref{proof1lemma12} and simplifying yields 
\vspace{-3pt}
\begin{equation}
\nonumber
  \delta_{l}=\mathbb{E}\left[\left(||\bm{g}_{l}||^{2}\right)^{2}\right]\hspace{-3pt}=
\hspace{-2pt}\mathbb{E}\big[\left(\bm{v}_{l}^{\textrm{H}}
  \bm{v}_{l}\right)^{2}\big]+
2\left(\bm{q}_{l}^{\textrm{H}}\bm{q}_{l}\right)
PM{\left(\eta'_{l}\right)}^{2}+2{\left(\eta'_{l}\right)}^{2}
\vspace{-3pt}
\end{equation}
\begin{equation}
\label{proof1lemma13}
\hspace{-140pt}
 \bm{q}_{l}^{\textrm{H}}\bm{A}_{l}\bm{A}_{l}^{\textrm{H}}\bm{q}_{l}
+\left(\bm{q}_{l}^{\textrm{H}}\bm{q}_{l}\bm{q}_{l}^{\textrm{H}}
\bm{q}_{l}\right). 
\vspace{-5pt}
\end{equation}
After recognizing that $\mathbb{E}\big[\hspace{-2pt}
\left(\bm{v}_{l}^{\textrm{H}}\bm{v}_{l}\right)^{2}\hspace{-2pt}\big]\hspace{-3pt}
=\hspace{-3pt}\mathbb{E}\left[\bm{v}_{l}^{\textrm{H}}\bm{v}_{l}\bm{v}_{l}^{\textrm{H}}
\bm{v}_{l}\right]$, substituting the definition of $\bm{v}_{l}$ and extracting the relevant 
constants allows us to write 
\vspace{-4pt}
\begin{equation}
\label{proof1lemma14}
\mathbb{E}\big[\left(
\bm{v}_{l}^{\textrm{H}}\bm{v}_{l}\right)^{2}
\hspace{-2pt}\big]\hspace{-1pt}
=\left(\eta'_{l}\right)^{\hspace{-2pt}4}\mathbb{E}\hspace{-2pt}\left[\left(
\bm{h}_{l}^{\textrm{H}}\bm{\Theta}\bm{h}_{l}\right)^{2}\right]\hspace{-3pt},
\vspace{-3pt}
\end{equation}
where $\bm{\Theta}=\bm{\Psi}^{\textrm{H}}\bm{\Gamma}\bm{\Psi}$ is an eigenvalue decomposition of 
$\bm{A}_{l}^{\textrm{H}}\bm{A}_{l}$. As a result,
\vspace{3pt}
\begin{equation}
\nonumber
\hspace{-51pt}
\mathbb{E}\left[\left(\bm{v}_{l}^{\textrm{H}}\bm{v}_{l}\right)^{2}\right]=
{\left(\eta'_{l}\right)}^{4}\hspace{2pt}\mathbb{E}\hspace{-2pt}\left[\left(\bm{h}_{l}^{\textrm{H}}
\bm{\Gamma}\bm{h}_{l}\right)^{2}\right]
\vspace{-5pt}
\end{equation}
\begin{equation}
\label{proof1lemma15}
\hspace{82pt}
={\left(\eta'_{l}\right)}^{4}\hspace{1pt}\mathbb{E}\left[\hspace{-1pt}\left(\sum\limits_{p=1}^{P}
\left[\bm{\Gamma}\right]_{p,p}|\bm{h}_{l;p}|^{2}\right)^{\hspace{-4pt}2}\hspace{2pt}\right], 
\vspace{-3pt}
\end{equation}
where $\bm{h}_{l;p}$ denotes the $p$-th element of $\bm{h}_{l}$. 
Performing the expectation with respect to $\bm{h}_{l}$ and further 
simplifying yields  
\vspace{-3pt}
\begin{equation}
\label{proof1lemma16}
\mathbb{E}\left[\left(\bm{v}_{l}^{\textrm{H}}\bm{v}_{l}\right)^{2}\right]
={\left(\eta'_{l}\right)}^{4}\hspace{2pt}\left\{
\left(\textrm{tr}\left[\bm{\Theta}\right]\right)^{2}+\textrm{tr}
\left[\bm{\Theta}^2\right]\right\}.
\vspace{-5pt}
\end{equation}
This allows us to write
\vspace{-2pt}
\begin{equation}
\label{proof1lemma17}
\mathbb{E}\left[\left(\bm{v}_{l}^{\textrm{H}}\bm{v}_{l}\right)^{2}\right]
\hspace{-1pt}=\hspace{-1pt}{\left(\eta'_{l}\right)}^{4}\left\{
\left(\textrm{tr}\left[
\bm{A}_{l}^{\textrm{H}}\bm{A}_{l}\right]\right)^{2}\hspace{-1pt}+
\hspace{-1pt}\textrm{tr}\left[\bm{A}_{l}^{\textrm{H}}\bm{A}_{l}
\bm{A}_{l}^{\textrm{H}}\bm{A}_{l}\right]\right\}\hspace{-1pt}. 
\vspace{-5pt}
\end{equation}
Recognizing that $\textrm{tr}\left[\bm{A}_{l}^{\textrm{H}}\bm{A}_{l}\right]=PM$ 
allows us to state
\vspace{-2pt}
\begin{equation}
\label{proof1lemma18}
\mathbb{E}\left[\left(\bm{v}_{l}^{\textrm{H}}\bm{v}_{l}\right)^{2}\right]=
{\left(\eta'_{l}\right)}^{4}\left\{P^{2}M^{2}+\textrm{tr}\left[\left(\bm{A}_{l}^{\textrm{H}}
\bm{A}_{l}\right)^{2}\right]\right\}. 
\end{equation}
Substituting the definition of $\bm{q}_{l}$ back, recognizing that 
$\mathbb{E}\left[\bar{\bm{h}}_{l}^{\textrm{H}}\bm{h}_{l}\right]=M$, 
combining \eqref{proof1lemma18} with the remaining terms in \eqref{proof1lemma13} 
and extracting the relevant constants results in the desired expression.

\vspace{-9pt}
\section{Proof of Lemma 2}
\label{ProofLemma2}
Applying the definition of $\bm{g}_{l}$ and $\bm{g}_{k}$ into 
$\mathbb{E}\left[|\bm{g}_{l}^{\textrm{H}}\bm{g}_{k}|^{2}\right]$ and denoting 
$\bm{v}_{l}=\eta'_{l}\bm{A}_{l}\bm{h}_{l}$ and $\bm{q}_{l}=\bar{\eta}_{l}\bar{\bm{h}}_{l}$ 
yields 
\begin{equation}
\label{proof2lemma21}
\varphi_{l}=\mathbb{E}\left[|\bm{g}_{l}^{\textrm{H}}\bm{g}_{k}|^{2}\right]
\hspace{-3pt}=\hspace{-3pt}\mathbb{E}\left[\big|\hspace{-2pt}\left(\bm{v}_{l}^{\textrm{H}}
\hspace{-2pt}+\hspace{-2pt}\bm{q}_{l}^{\textrm{H}}\right)
\left(\bm{v}_{k}\hspace{-2pt}+\hspace{-2pt}\bm{q}_{k}\right)\big|^{2}\right]. 
\vspace{-5pt}
\end{equation}
Expanding and simplifying \eqref{proof2lemma21} allows us to state 
\vspace{-2pt}
\begin{equation}
\nonumber
\varphi_{l}=\mathbb{E}\left[|\bm{g}_{l}^{\textrm{H}}\bm{g}_{k}|^{2}\right]=
\mathbb{E}\left[\left(\bm{v}_{l}^{\textrm{H}}\bm{v}_{k}+\bm{v}_{l}^{\textrm{H}}
\bm{q}_{k}+\bm{q}_{l}^{\textrm{H}}\bm{v}_{k}+\bm{q}_{l}^{\textrm{H}}\bm{q}_{k}
\right)\right.
\vspace{-2pt}
\end{equation}
\begin{equation}
\label{proof2lemma22}
\hspace{-105pt}
\left.\left(\bm{v}_{k}^{\textrm{H}}\bm{v}_{l}+\bm{q}_{k}^{\textrm{H}}\bm{v}_{l}
+\bm{v}_{k}^{\textrm{H}}\bm{q}_{l}+\bm{q}_{k}^{\textrm{H}}\bm{q}_{l}\right)\right].
\vspace{-2pt}
\end{equation}
Further expanding and simplifying yields 
\vspace{-5pt}
\begin{equation}
\nonumber
\varphi_{l}=\mathbb{E}\left[|\bm{g}_{l}^{\textrm{H}}\bm{g}_{k}|^{2}\right]=
\mathbb{E}\left[\bm{v}_{l}^{\textrm{H}}\bm{v}_{k}\bm{v}_{k}^{\textrm{H}}\bm{v}_{l}\right]
+\mathbb{E}\left[\bm{v}_{l}^{\textrm{H}}\bm{q}_{k}\bm{q}_{k}^{\textrm{H}}\bm{v}_{l}\right]+
\vspace{-1pt}
\end{equation}
\begin{equation}
\label{proof2lemma23}
\hspace{-95pt}
\mathbb{E}\left[\bm{q}_{l}^{\textrm{H}}\bm{v}_{k}\bm{v}_{k}^{\textrm{H}}\bm{q}_{l}\right]+
\mathbb{E}\left[\bm{q}_{l}^{\textrm{H}}\bm{q}_{k}\bm{q}_{k}^{\textrm{H}}\bm{q}_{l}\right]. 
\vspace{-3pt}
\end{equation}
Invoking the independence between $\bm{v}_{l}$ and $\bm{v}_{k}$, recognizing 
that $\mathbb{E}\left[\bm{v}_{l}\bm{v}_{l}^{\textrm{H}}\right]
=\left(\eta'_{l}\right)^{2}\textrm{tr}\left[\bm{A}_{l}^{\textrm{H}}\bm{A}_{l}\right]$, 
upon substituting back the definitions of $\bm{v}_{k}$ and $\bm{q}_{k}$ and extracting the 
relevant constants, we can state 
\vspace{-4pt}
\begin{equation}
\nonumber
\varphi_{l}=\mathbb{E}\left[|\bm{g}_{l}^{\textrm{H}}\bm{g}_{k}|^{2}\right]
\hspace{-3pt}=\hspace{-3pt}
\left(\eta'_{l}\right)^{\hspace{-1pt}2}\hspace{-2pt}\left(\eta'_{k}\right)^{\hspace{-1pt}2}
\mathbb{E}\left[\bm{h}_{l}^{\textrm{H}}\hspace{-2pt}\bm{A}_{l}^{\textrm{H}}
\hspace{-2pt}\bm{A}_{k}\hspace{-2pt}\bm{A}_{k}^{\textrm{H}}\right]\hspace{-2pt}+
\hspace{-2pt}\left(\eta'_{l}\right)^{2}\hspace{-2pt}\left(\bar{\eta}_{k}\right)^{2}
\left[\bar{\bm{h}}_{k}^{\textrm{H}}\hspace{-2pt}\bm{A}_{l}\right.
\vspace{-3pt}
\end{equation}
\begin{equation}
\label{proof2lemma24}
\hspace{-3pt}
\left.
\hspace{-1pt}\bm{A}_{l}^{\textrm{H}}\hspace{-1pt}
\bar{\bm{h}}_{k}\right]\hspace{-3pt}+\hspace{-3pt}\left(\bar{\eta}_{l}\right)^{2}
\hspace{-3pt}\left(\eta'_{k}\right)^{2}\hspace{-2pt}
\left[\bar{\bm{h}}_{l}^{\textrm{H}}\hspace{-1pt}\bm{A}_{k}\hspace{-1pt}
\bm{A}_{k}^{\textrm{H}}\hspace{-1pt}
\bar{\bm{h}}_{l}\right]\hspace{-3pt}+\hspace{-3pt}
\left(\bar{\eta}_{l}\right)^{2}\hspace{-2pt}
\left(\bar{\eta}_{k}\right)^{2}\left[|\bar{\bm{h}}_{l}^{\textrm{H}}
\hspace{-1pt}\bar{\bm{h}}_{k}|^{2}\right]. 
\vspace{-3pt}
\end{equation}
Taking the trace and simplifying yields 
the result in \eqref{lemma2}. 

\vspace{-10pt}
\section*{Acknowledgment}
\label{Acknowledgment}
The authors would like to thank Prof. Andreas F. Molisch at 
the University of Southern California for the insightful 
discussions during the course of this work. 

\vspace{-4pt}
\bibliographystyle{IEEEtran}

\begin{thebibliography}{1}
\bibitem{RUSEK} F. Rusek, D. Persson, B. Lau, E. G. Larsson, T. L. Marzetta, O.~Edfors, and F.~Tufvesson, ``Scaling up MIMO: Opportunities and challenges with very large arrays", \emph{IEEE Signal Process. Mag.}, vol. 30, no. 1, pp. 40-60, Nov. 2013. 

\bibitem{LARSSON} E. G. Larsson, O. Edfors, F. Tufvesson, and T. L. Marzetta, ``Massive MIMO for next generation wireless systems", \emph{IEEE Commun. Mag.}, vol. 52, no. 2, pp. 186-195, Feb. 2014. 

\bibitem{MASOUROS1} C. Masouros, M. Sellathurai, and T. Ratnarajah, ``Large-scale MIMO transmitters in fixed physical spaces: The effect of transmit correlation and mutual coupling", \emph{IEEE Trans. Commun.}, vol. 61, no. 7, pp. 2794-2804, Jul. 2013. 

\bibitem{MASOUROS2} C. Masouros and M. Matthaiou, ``Space-constrained massive MIMO: Hitting the wall of favorable propagation", \emph{IEEE Commun. Lett.}, vol. 19, no. 5, pp. 771-774, May 2015.

\bibitem{NGO} H. Q. Ngo, E. G. Larsson, and T. L. Marzetta, ``The multicell multiuser 
MIMO uplink with very large antenna arrays and a finite-dimensional channel", 
\emph{IEEE Trans. Commun.}, vol. 6, no. 61, pp. 2350-2361, Jun. 2013. 


\bibitem{GARCIA-ROD} A. Garcia-Rodriguez and C. Masouros, ``Exploiting the increasing correlation of space constrained massive MIMO for CSI relaxation", 
\emph{IEEE Trans. Commun.}, vol. 4, no. 64, pp. 1572-1587, Apr. 2016. 


\bibitem{ZHANG} J. Zhang, L. Dai, M. Matthaiou, C. Masouros, and S. Jin, ``On the spectral efficiency of space-constrained massive MIMO with linear receivers", in 
\emph{Proc. of IEEE Int. Conf. on Commun. (ICC)}, pp. 1-6, May 2016. 


\bibitem{BISWAS} S. Biswas, C. Masouros, and T. Ratnarajah, ``Performance analysis of large multi-user MIMO systems with space-constrained 2D antenna arrays", 
\emph{IEEE Trans. Wireless Commun.}, vol. 15, no. 5, pp. 3492-3505, May 2016. 

 
\bibitem{GE} X. Ge, R. Zi, H. Wang, J. Zhang, and M. Jo, ``Multi-user massive MIMO communication systems based on irregular antenna arrays", \emph{IEEE Trans. Wireless Commun.}, vol. 15, no. 8, pp. 5287-5301, Aug. 2016.


\bibitem{AKDNENIZ} M. R. Akdeniz, Y. Liu, M. K. Samimi, S. Sun, S. Rangan, T. S. Rappaport, and E.~Erkip, ``Millimeter wave channel modeling and cellular capacity evaluation", \emph{IEEE J. Sel. Areas Commun.}, vol. 32, no. 6, pp. 1164-1179, Jun. 2014.


\bibitem{SUN} S. Sun, T. S. Rappaport, R. W. Heath Jr., A. R. Nix, and S. Rangan, ``MIMO for millimeter-wave wireless communications: Beamforming, spatial multiplexing, or both?", \emph{IEEE Commun. Mag.}, vol. 52, no. 12, pp. 110-121, Dec. 2014. 


\bibitem{TATARIA} H. Tataria, P. J. Smith, L. J. Greenstein, P. A. Dmochowski, and M. Shafi, ``Performance and analysis of downlink multiuser MIMO systems with regularized zero-forcing precoding in Ricean fading channels", in \emph{Proc. of IEEE Int. Conf. on Commun. (ICC)}, pp. 1185-1192, May 2016. 


\bibitem{TATARIA2} H. Tataria, P. J. Smith, L. J. Greenstein, and P. A. Dmochowski, 
``Zero-forcing precoding performance in multiuser MIMO systems with heterogeneous Ricean fading", \emph{IEEE Wireless Commun. Lett.}, vol. 6, no. 1, pp. 74-77, Feb. 2017. 


\bibitem{NAM} J. Nam, G. Caire, and J. Ha, ``On the role of transmit correlation diversity in multiuser MIMO systems", \emph{IEEE Trans. Info. Theory}, vol. 63, no. 1, pp. 336-354, Jan. 2017. 


\bibitem{MOLISCH} A. F. Molisch \emph{Wireless Communications}, Wiley Press, 2011. 


\bibitem{WARNICK} K. Warnick and M. Jensen, ``Optimal noise matching for mutually coupled arrays", \emph{IEEE Trans. Antennas Propag.}, vol. 55, no. 6, pp. 1726-1731, Jun. 2007.


\bibitem{3GPPTR36873} 3GPP TR 36.873 v12.0.0, \emph{Study on 3D channel models for LTE}, 
3GPP, Jun. 2015.


\bibitem{ZHANG2} Q. Zhang, S. Jin, K-K. Wong, H. Zhu, and M. Matthaiou, ``Power scaling of uplink massive MIMO systems with arbitrary-rank channel means", \emph{IEEE J. Sel. Topics Signal Process.}, vol. 8, no. 5, pp. 966-981, Nov. 2014. 


\bibitem{BASNAYAKA} D. Basnayaka, P. J.~Smith, and P. A.~Martin, ``Performance analysis of macrodiversity MIMO systems with MMSE and ZF receivers in flat Rayleigh fading", \emph{IEEE Trans. Wireless Commun.}, vol. 12, no. 5, pp. 2240-2251, May 2013. 


\bibitem{THOMAS} T. Thomas, H. C. Nguyen, G. R. MacCartney, and T. S. Rappaport, 
``3D mmWave channel model proposal", in \emph{Proc. IEEE Conf. on Veh. Technol. (VTC-Fall)}, pp. 1-6, Sep. 2014. 
\end{thebibliography}

\end{document}